\documentclass[default,iicol]{sn-jnl}

\usepackage{float}
\usepackage{listings}
\usepackage{hyperref}

\lstdefinelanguage{yaml}{
     keywordstyle=\color{blue}\ttfamily,  
     keywordsprefix=Tuple,
     morekeywords={TupleToolKinematic,TupleToolPid,TupleToolANNPID, TupleToolGeometry,TupleToolEventInfo,TupleToolTISTOS},
     basicstyle=\color{blue}\ttfamily,
     rulecolor=\color{black},
     string=[s]{'}{'},
     stringstyle=\color{red},
     comment=[l]{:},
     commentstyle=\color{black}\ttfamily,
     morecomment=[l]{-},
     morecomment=[l]{Hlt1(},
     morecomment=[l]{Hlt2(},     
     morecomment=[l]{(?!},
     breaklines=true,
 }

\def\wizard{Ntuple Wizard}

\jyear{2021}%

\theoremstyle{thmstyleone}%
\theoremstyle{thmstyletwo}%

\theoremstyle{thmstylethree}%

\begin{document}

\title[The LHCb Ntuple Wizard]{Ntuple Wizard: An Application to Access Large-Scale Open Data from LHCb}

\author[1]{\fnm{Christine A.} \sur{Aidala}}

\author[2]{\fnm{Christopher} \sur{Burr}}

\author[2]{\fnm{Marco} \sur{Cattaneo}}

\author*[1]{\fnm{Dillon S.} \sur{Fitzgerald}}\email{dillfitz@umich.edu}

\author[2,3]{\fnm{Adam} \sur{Morris}}

\author[3]{\fnm{Sebastian} \sur{Neubert}}

\author[2,4]{\fnm{Donijor} \sur{Tropmann}}

\affil*[1]{\orgdiv{Department of Physics}, \orgname{University of Michigan}, \orgaddress{\street{450 Church St}, \city{Ann Arbor}, \postcode{48109}, \state{Michigan}, \country{USA}}}

\affil[2]{\orgname{European Organization for Nuclear Research (CERN)}, \orgaddress{\city{Geneva},  \country{Switzerland}}}

\affil[3]{\orgname{Universit\"at Bonn -- Helmholtz-Institut f\"ur Strahlen und Kernphysik}, \orgaddress{\city{Bonn}, \country{Germany}}}

\affil[4]{\orgname{RWTH Aachen University}, \orgaddress{\city{Aachen}, \country{Germany}}}


\abstract{Making the large data sets collected at the Large Hadron Collider (LHC) accessible to the world is a considerable challenge because of both the complexity and the volume of data. This paper presents the Ntuple Wizard, an application that leverages the existing computing infrastructure available to the LHCb collaboration in order to enable third-party users to request specific data. An intuitive web interface allows the discovery of accessible data sets and guides the user through the process of specifying a configuration-based request. The application allows for fine-grained control of the level of access granted to the public.  }




\maketitle

\section{Introduction}
\label{sec:Introduction}

In an increasingly diverse research landscape, management and curation of public data are becoming critical components of transdisciplinary science. Keys to the realization of an open research ecosystem that adds scientific value have been identified in the FAIR principles of scientific data management and stewardship \cite{fair_2016}. Making data Findable, Accessible, Interoperable, and Reusable, however, requires a considerable amount of tooling and infrastructure.

A common problem, which is acute for data in high-energy physics but increasingly an issue in other fields as well, is the sheer size of data sets stored in custom file formats.
For large-scale experimental facilities, such as the Large Hadron Collider (LHC) at the European Organization for Nuclear Research (CERN), the data sets are so large that even access by the directly involved scientists has to be centrally managed. LHCb is one of the four major experiments at the LHC~\cite{LHCb_det}, and as an example, the data collected by the LHCb experiment in the years 2011-12 (corresponding to $\sim 3$ fb$^{-1}$ of proton-proton collisions) amount to a volume of 900 TB of reconstructed data. For the purpose of processing these data, extensive computing infrastructure has been set up by the countries participating in this type of research \cite{Bird:2014ctt}. Making use of this infrastructure to enable external users FAIR access to LHCb data in a resource-effective way is the goal of the application presented in this paper. 

In the following, an application is presented that exposes a data query service to allow the public to request sub-samples of data collected and published by the LHCb experiment. The samples are delivered as ROOT Ntuples~\cite{ROOT} a data format that requires no special LHCb-specific software to read and for which converters to other standard file formats exist. We call the application the \wizard{}.

The Ntuple Wizard interface guides users with basic knowledge in particle physics through the process of discovering the available data and formulating a useful query. The queries can be processed by the existing data production infrastructure, and results will be delivered through the CERN Open Data Portal~\cite{ODP}. By splitting the data request into the construction of a data query and subsequent processing of the query on the internal infrastructure, the LHCb collaboration retains fine-grained control over access to the data. Crucially this system protects the compute infrastructure from attacks by malicious code injection.

\subsection{Accessible open data}
In 2020, the LHC experiments at CERN adopted a new Open Data Policy~\cite{CERN-OPEN-2020-013}, the scope of which expanded in 2022 to an Open Science Policy~\cite{CERN-OPEN-2022-013}. These documents define the commitments of CERN to make the data collected at the LHC, at several levels of complexity, publicly available~\cite{DPHEPStudyGroup:2012dsv}:

\begin{description}
\item[Level 1] Published results --- this can include tables and figures but also preprocessed Ntuples or binned and unbinned fit likelihood functions. 
\item[Level 2] Outreach and education --- usually in the form of highly preprocessed Ntuples.
\item[Level 3] Reconstructed data --- these data have been preprocessed to derive physics objects, such as charged particle candidates, photons, or particle jets. Reconstructed data may or may not be corrected for detector effects, such as efficiency and resolution.
\item[Level 4] Raw data -- the basic quantities recorded by the experimental instruments.
\end{description}

Both Level 1 and 2 data are considered to be highly processed, abstracted, and manageable using commonly available computers. Level 4 raw data will not be made available due to practical reasons concerning data size but also detector-specific information needed for the interpretation of these data. This leaves Level 3 data as the most versatile and basic data set which will be publicly accessible.

All LHC experiments have long and intricate data reconstruction pipelines, which yield several intermediate output data formats. During a pipeline, the raw data are converted to physical objects such as charged particle trajectories, jets, and vertices. Furthermore, the raw data are classified and filtered to obtain samples enriched in interesting signatures. 

\begin{figure*}[!htbp]
    \centering
    \includegraphics[width=\textwidth]{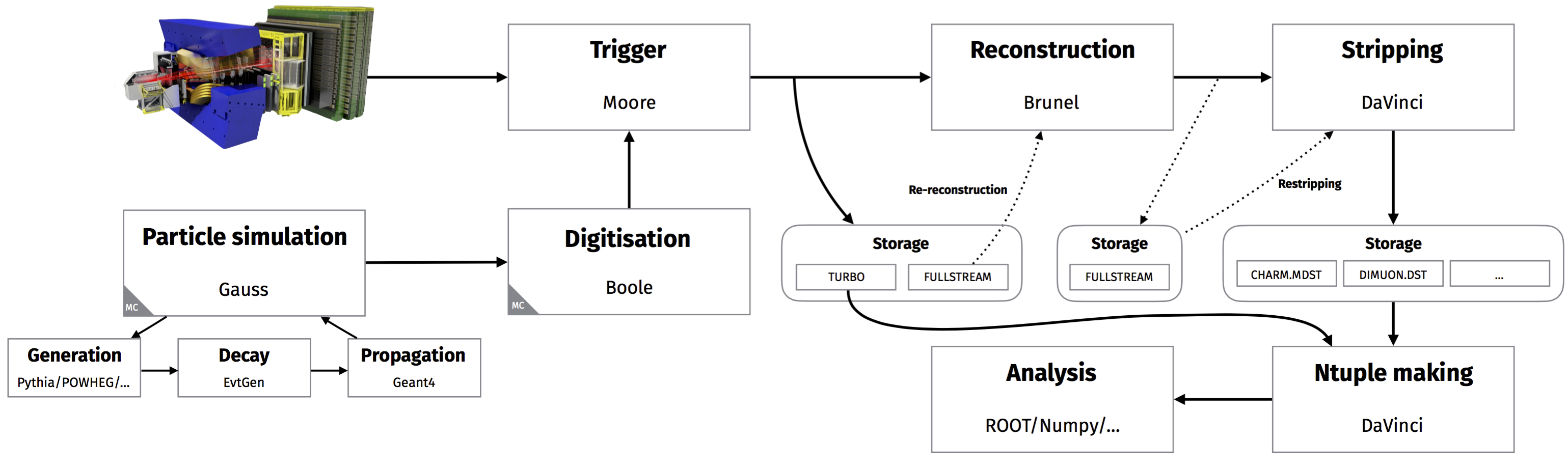}
    \caption{LHCb data flow in Runs 1 and 2. The output of the stripping step will be made public through the CERN Open Data Portal~\cite{ODP}.}
    \label{fig:data_flow_run1and2}
\end{figure*}

Figure \ref{fig:data_flow_run1and2} shows an overview of the data processing pipeline in LHCb as it has been used during LHC data-taking Runs 1 and 2 (2011--18). The various steps of the pipeline are outlined further in the LHCb computing technical design reports~\cite{TDR1, TDR2}. Level 3 data have been defined as the output of the {\it stripping} step. The stripping consists of a large number of selection algorithms called {\it lines}, which are designed to filter the data and sort events into several collections, which are called {\it streams}. Streams are defined according to common physics signatures and aim to collect selections with significant overlaps into a common set of files, to reduce duplication of the data. The LHCb data organization is discussed in more detail in Appendix~\ref{app:data_org}, including a list of streams available in Runs 1 and 2.

The stripping selections are based on the concept of physics {\it candidates}. A candidate refers to a set of data matching a particular physics signature. In most cases, this signature will be a particular particle decay, such as for example $B^+ \rightarrow \bar{D}^0 \pi^+$ with the subsequent decay $\bar{D}^0 \rightarrow K^+\pi^-$, where $B, D, K,$ and $\pi$ mesons are the lightest hadrons containing $b, c, s$, and $u/d$ quarks respectively. Such cascading decays are represented as tree-like data structures, where the nodes represent (intermediate) particles and the edges indicate a parent-child relationship in the decay. These data structures are referred to as {\it decay trees} (see Fig.~\ref{fig:node_tree} for an example). The root particle of the decay tree (the $B^+$ in our example) is called its {\it head}. Stripping selections attempt to find sets of physics objects in the reconstructed LHCb data, which match the desired decay tree and any additional criteria that might be applied to distinguish the intended signal process from background. Typical selection criteria include kinematic variables, vertex and track reconstruction qualities, and particle identification variables. Some stripping lines rely on multivariate classifiers to combine several observables into a single powerful selection criterion. The output of this procedure is collections of decay candidates specified by their particular decay trees in a custom LHCb-specific data format. 

It is important to note that candidates are distinct from the concept of {\it events} in the LHCb data processing. An event is defined during the data acquisition and refers to a particular time window in which collisions can occur. Several proton-proton collisions can happen during this time window, and in principle, it can happen that several candidates for a particular decay are identified for a single collision. In such cases, relevant quantities (related to vertex reconstruction and flight distances) can be computed for every primary vertex (e.g. collision point) in the event.

In order to convert these data into a framework-independent format a useful concept is the aforementioned Ntuples. The idea of a Ntuple is simple: each candidate is described by a tuple of variables, i.e. physical observables of interest measured on the particular candidate, or referring to the global event in which the candidate was found. A data set consists of $N$ such tuples, much like a simple CSV file. Ntuples are saved in ROOT files~\cite{ROOT} and only basic data types are allowed for the variables. As a small complication in some instances, the variables can actually be arrays of basic data types. In such cases, the \wizard{} provides the necessary documentation for the interpretation.

\subsection{Principle of Ntuple creation and the \wizard{}}
Both the stripping as well as the Ntuple-making step in Fig.~\ref{fig:data_flow_run1and2} are handled by DaVinci~\cite{DaVinci, TDR1, TDR2}, an LHCb application for event selection and data analysis using the Gaudi framework~\cite{Gaudi, TDR1, TDR2}. DaVinci is configured via Python scripts and used to process entire data sets with batch processing. Both the Python configuration as well as the batch production system are intentionally hidden from users of the \wizard{} for security reasons.


The DaVinci application provides access to a number of algorithms that can be combined in sequence for event selection and processing. In order to produce a Ntuple the user has to specify which variables should appear in the output data. This Ntuple configuration is handled by an algorithm named {\it DecayTreeTuple}, in which variables are registered through the use of so-called  {\it TupleTools} and {\it LoKi functors}. A large collection of those tools and functors are available for the user to choose from. In general, a TupleTool will add a set of variables to the Ntuple, while a LoKi functor usually computes a single number. The {\it LoKi::Hybrid::TupleTool} can be used to write the output of functors into the tuple. Functors can be combined with standard arithmetic and logic operations, providing a flexible and powerful system to compute derived quantities. A list of important available tools is presented in Appendix~\ref{app:tupletools}.


\begin{figure*}[!htbp]
    \centering
    \includegraphics[width=\textwidth]{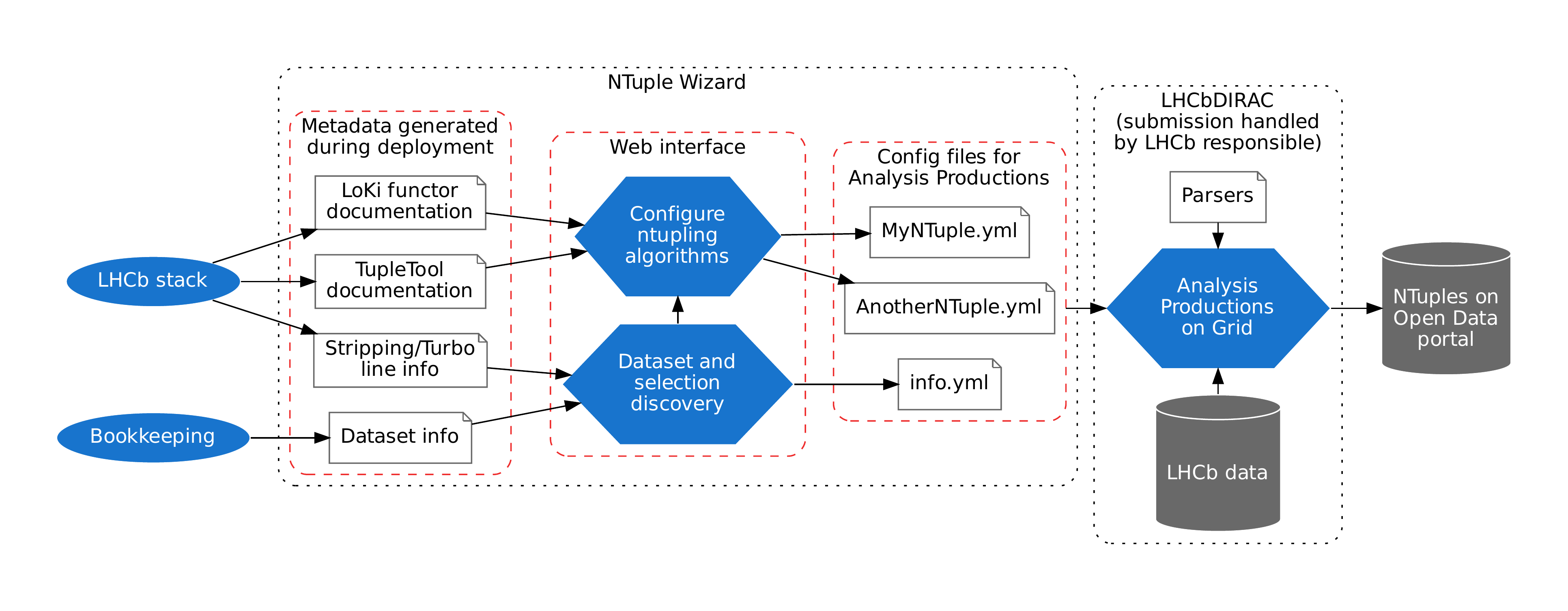}
    \caption{Architecture of the \wizard{}.}
    \label{fig:arch}
\end{figure*}

Figure \ref{fig:arch} shows an overview of the \wizard{} architecture, the core functionality of which is the configuration of DaVinci. The metadata and documentation describing the available data, preselections, as well as available selection operations are generated from the original provenance traces of the data and the stripping selection code. The web interface presents this metadata and documentation to the user in a pedagogical way that facilitates data discovery and formulation of the query. The query to the data has two principal parts: Data set discovery and Ntuple configuration. First, the Ntuple Wizard allows the user to select from the available predefined stripping selections, data-taking periods, and conditions. In the second step, the user defines what quantities should be computed and written to the output Ntuple. Standard tools for the computation of typical quantities, such as kinematic variables, particle identification (PID) variables, etc., are available. The query formulated by the user is stored in a set of configuration files. These files can be converted into a Python configuration compatible with the internal LHCb Analysis Productions system~\cite{AnalysisProductions}. This conversion and the final submission of the query to the computing infrastructure are handled through an LHCb Analysis Productions manager.

\subsection{Security considerations}
Accepting arbitrary external code to run on the LHCb computing resources has obvious unacceptable security risks. Therefore, the \wizard{} is designed to generate the configuration in a pure data-structure format. As shown in Figure~\ref{fig:arch}, the configuration of the query is captured in YAML files, which can be submitted to an LHCb Analysis Productions manager for further processing.

\section{Metadata and documentation acquisition}
\label{sec:meta}

In order to facilitate the core functionality of the \wizard{} --- namely data set discovery (Sec.~\ref{sec:discover}) and algorithm configuration (Sec.~\ref{sec:alg}), metadata and documentation from several sources are required. In particular, the Ntuple Wizard needs to know what types of decays can be queried and what tools are available to compute derived quantities of interest about these candidates.

Since these metadata are unchanging, and providing direct access to the various sources requires authentication and introduces more points of failure, the metadata are collated and served as static files over HTTP. No additional access to the LHCb code or database is needed by the \wizard{} once it has been deployed.

The sources of metadata can be grouped into two coarse categories: the LHCb software stack and the LHCb database. Metadata are acquired from the LHCb software stack in two ways. The first is from the Gaudi Python interface; particularly under the DaVinci application environment. Metadata about the configuration interface of each TupleTool are extracted from DaVinci. Details of the stripping lines, including the chain of selection algorithms that define them, are extracted from the DaVinci versions used in the corresponding stripping campaigns.

The process of building decay candidates in a stripping line often involves a combination of many algorithms from the LHCb selection framework, which combine particles, impose selection requirements, perform PID substitution, and build final-state particle candidates from trajectories of charged particles and calorimeter clusters.
The algorithms can be related to each other through their input and output locations. The full list of decays (including all sub-decays) must be inferred by traversing the `dependency tree' of the selection algorithms. This is performed using custom code during metadata acquisition.

The second, more indirect way is from the LHCb Doxygen pages, which themselves are generated from the source code of the LHCb software stack.
The latest Doxygen pages for Run 1 or Run 2 DaVinci versions are used to extract the documentation for each TupleTool and LoKi functor.
A campaign to improve the Doxygen documentation at its source was undertaken during the development of the \wizard{}.

The LHCb database provides metadata about the centrally managed data sets, which is necessary to configure the Ntupling productions as explained above.
In order not to duplicate effort, a common code base is employed to extract metadata from the LHCb database for both the \wizard{} and the CERN Open Data Portal.

\section{User interface}
\label{sec:web}

The user interface consists of a sequence of dialogues that guide the user through the configuration steps. This is designed as a client-side dynamic web page that reads metadata acquired at deployment time to serve as static files (see Sec.~\ref{sec:meta}).

Since users of LHCb open data do not, in general, have access to the same support network of experienced users and developers enjoyed by LHCb collaboration members, a key design element of the Wizard is to provide the necessary documentation for a novice user to complete each step of the configuration process.


The existing documentation of DaVinci~\cite{DaVinci, TDR1, TDR2} is fragmented across several sources (Twiki~\cite{twiki}, the Starterkit~\cite{starterkit}, Doxygen~\cite{doxygen} and the source code itself), so where possible, the Wizard pulls text from each of these disparate sources and renders it in the relevant context within the user interface.

There are two main steps to formulate a query using the \wizard{}: Data set discovery and Ntuple configuration. These steps are explained in the following.

\section{Data set discovery and production configuration}
\label{sec:discover}

The available data contain a wide range of stripping selections, data-taking periods, and running conditions. The {\bf Production configuration} dialogue of the \wizard{} guides the user through the selection of the desired subsets. The interface allows the selection of several decays to be processed simultaneously as part of one query. For each decay, a separate Ntuple will be produced. 

\subsection{Discovering available candidate decays}

\begin{figure*}[!htbp]
    \centering
    \includegraphics[width=\textwidth]{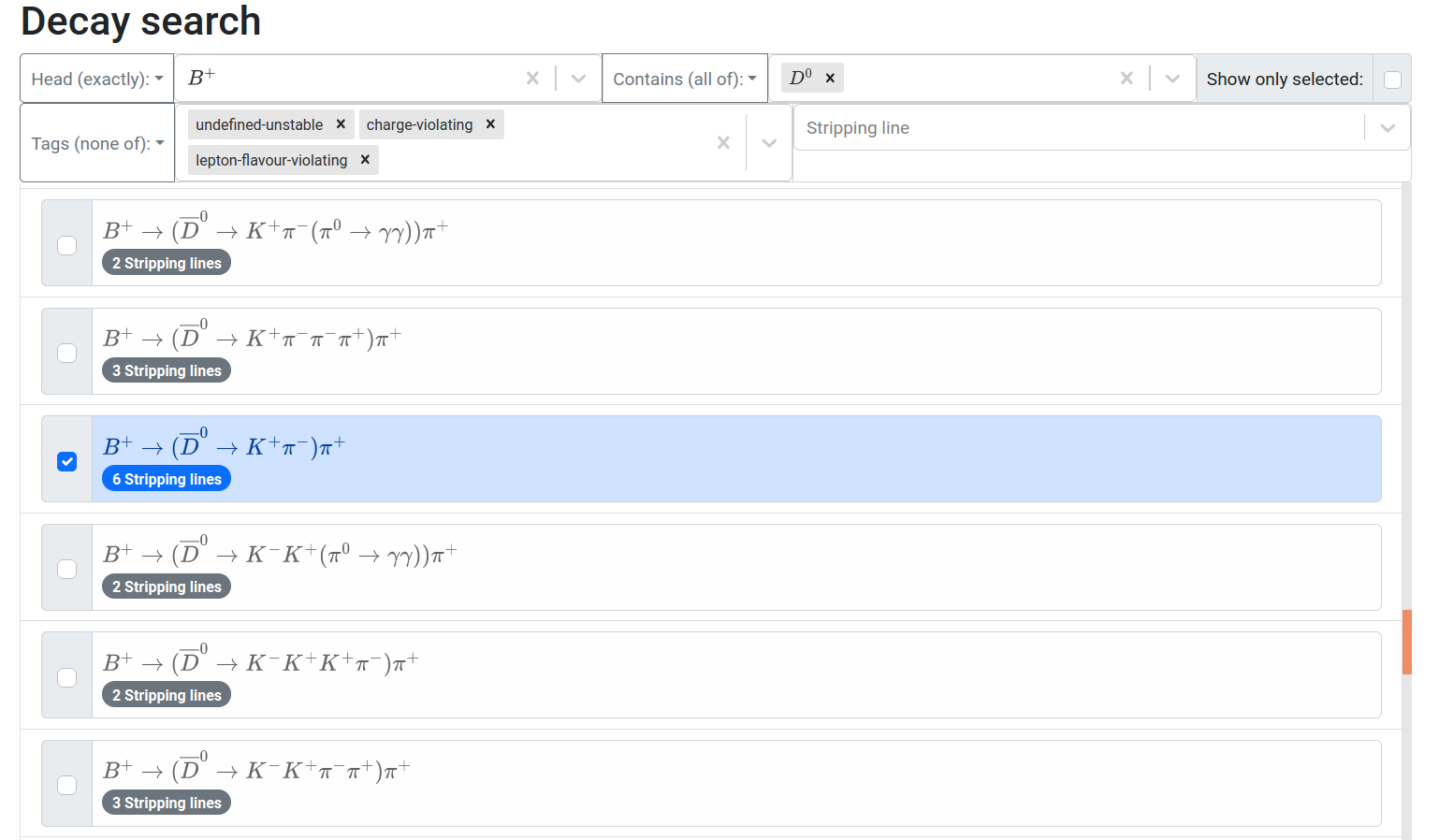}
    \caption{Example of the decay candidate search function of the \wizard{}.}
    \label{fig:decay_search}
\end{figure*}

In the {\bf Decay search} dialogue, the \wizard{} presents a list of all decays selected by the stripping, accompanied by decay descriptors in LoKi and LaTeX formats, information about which stripping lines build them, as well as `tags' that can be used to filter different types of decays. Decays are searchable through various filters, including the identity or properties of the parent particle and decay products, whether the candidates are built by a specific stripping line, and the aforementioned tags. An example of the decay search is shown in Figure~\ref{fig:decay_search}. The selected candidate of interest is highlighted in blue, and the collection was narrowed down from the list of all possible decays by using the filters and tags at the top of the page. The `none of' option of the `tags' drop-down menu is chosen by default, indicating that decays with the displayed tags are hidden from the list of selectable decays. The tags `charge-violating` and `undefined-unstable` corresponding to decays that violate charge conservation and that contain unstable particles without defined decays respectively are hidden by default. If the user wishes to instead isolate decays that meet the criteria of any or all of the selected tags, the options `any' or `all' can be chosen from the `tags' drop-down menu. It is possible to select several decays for further processing at this stage.   

\subsection{Stripping line and data set selection}

Once a decay is selected by the user, all corresponding stripping lines and data sets from the various running periods are listed, and the desired combination(s) can be selected. The case can arise where the same stripping line shows up in multiple stripping versions within the same data set (stream, running year, and magnet polarity). These are rendered as separate options in the data set selection drop-down menu of the \wizard{}. For a given decay, it is recommended to choose only one data set for each magnet polarity within a given running year, and to use the most recent stripping version in the case of duplicates. The data organization of LHCb is elaborated on in Appendix~\ref{app:data_org}, including a table of running years, as well as corresponding collision energies and stripping versions.  

\begin{figure*}[!htbp]
    \centering
    \includegraphics[width=\textwidth]{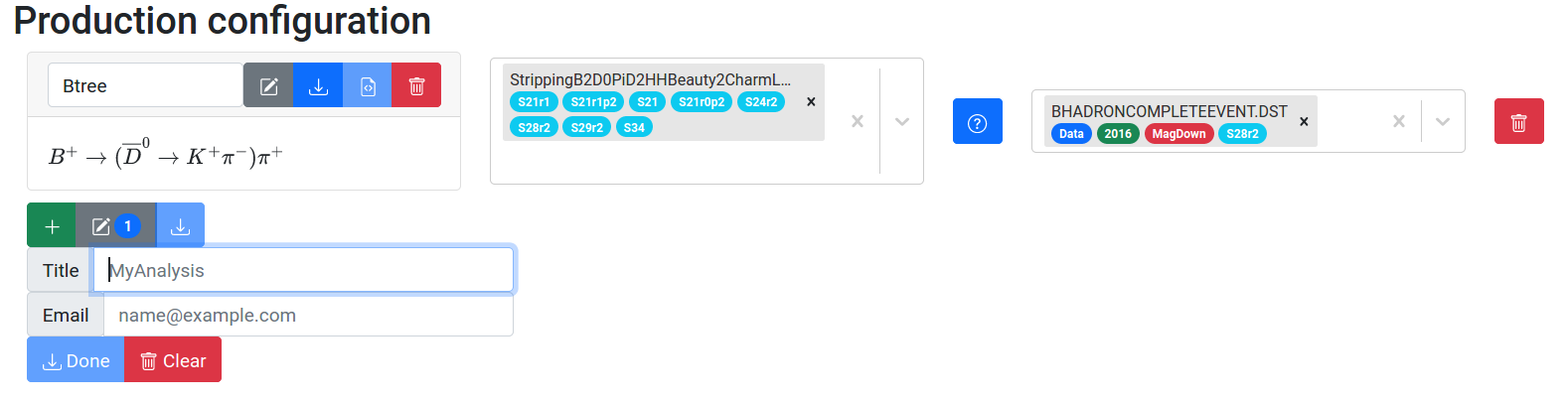}
    \caption{Example of the data set selection and production configuration step of the \wizard{}.}
    \label{fig:prod_config}
\end{figure*}

Links to documentation about each stripping line including selection algorithms that went into building the decay candidates are displayed to the user to guide them in choosing the most suitable stripping line and data stream for their physics interest. Figure~\ref{fig:prod_config} shows an example of the production configuration page, where an available stripping line and data set have been chosen from lists of all possibilities corresponding to the selected decay channel. The blue question mark button contains links to the aforementioned stripping documentation. At this point, the query is specified up to deciding what information to write into the Ntuple.

\section{Ntuple configuration}
\label{sec:alg}
The {\bf DecayTreeTuple configuration} dialogue is designed to guide the user through customization of the quantities written to the Ntuple for the selected candidates. For each decay, a separate DecayTreeTuple has to be configured. Care should be taken to name the Ntuples appropriately. The \wizard{} requires a unique name for each Ntuple.
 
 \begin{figure*}[!htbp]
    \centering
    \includegraphics[width=\textwidth, height=0.55\textheight]{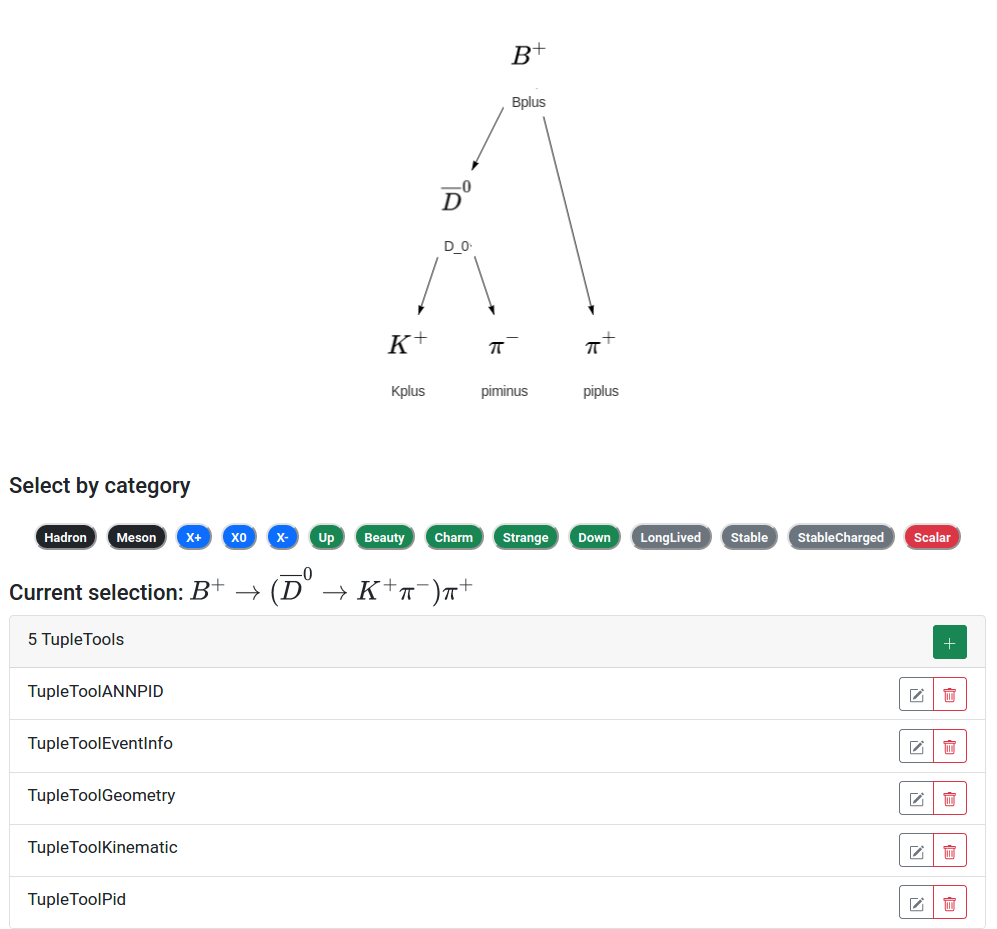} \\
    \vspace{0.75em}
    \includegraphics[width=\textwidth, height=0.4\textheight]{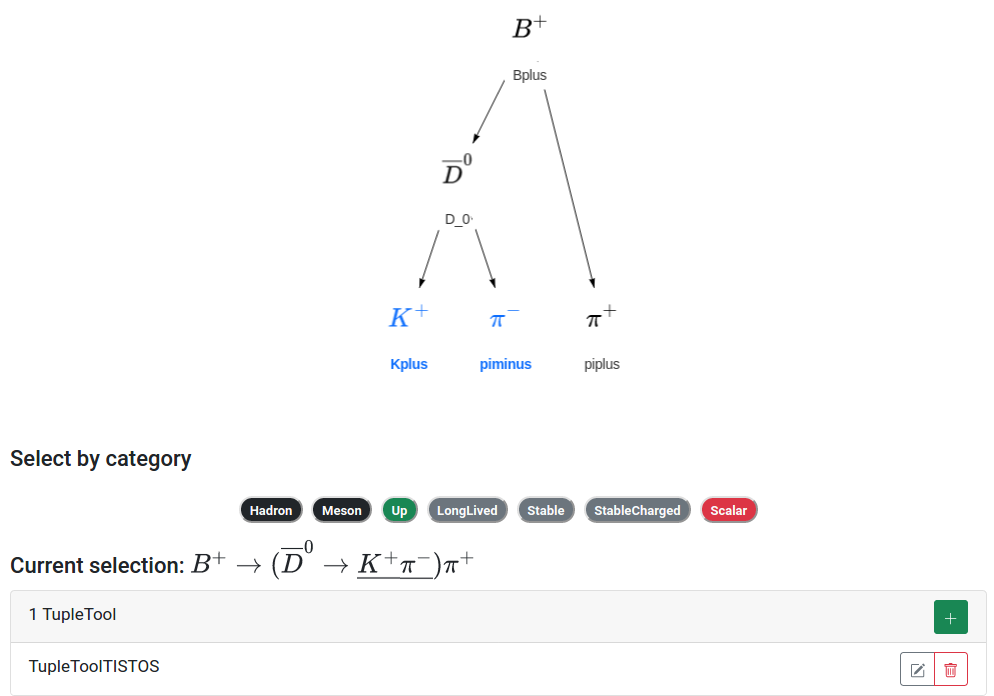}
    \caption{Example of an interactive graph used to configure DecayTreeTuple, with selected TupleTools displayed for both the entire candidate (top) and selected nodes (bottom).}
    \label{fig:node_tree}
\end{figure*}

 Selected decay trees are visually represented as graphs, where each physics object (e.g. particle) is represented by a node as shown in the screenshots in Figure \ref{fig:node_tree}. The user can interact with this graph by selecting one or multiple nodes at a time and determining which TupleTools will be added to each node, which in turn determines which quantities are saved to the Ntuple. A list is rendered on screen depending on the selected node(s), each element of which corresponds to a selected TupleTool, with buttons for configuring and removing the tool. The TupleTool configuration interface includes links to relevant documentation about the tool, including lists of quantities written by the tool where available. Each node in the graph comes with the standard set of TupleTools for LHCb analyses, but more will often be needed depending on the particular physics interests of the user.  Furthermore, any added tool will come with the standard configuration, which can be further modified if the user desires. A custom set of standard LoKi variables and functions of these variables can also be saved to the Ntuple for each node, using the {\it Loki::Hybrid::TupleTool}. Appendix~\ref{app:tupletools} contains a brief description of the standard set of TupleTools included with each node on the graph, as well as other useful TupleTools for physics analysis. Figure~\ref{fig:node_tree} shows an example of the configurable graph corresponding to the selected candidate shown in Figures~\ref{fig:decay_search} and~\ref{fig:prod_config}, as well as a list of TupleTools corresponding to the entire decay candidate (top), and particular nodes selected on the graph (bottom). It can be seen from the figure that nodes can also be selected through the categories shown below the graph and that TupleTools can be added, removed, or configured for each node or grouping of nodes.

\begin{figure*}[!htbp]
    \centering
    \includegraphics[width=\textwidth]{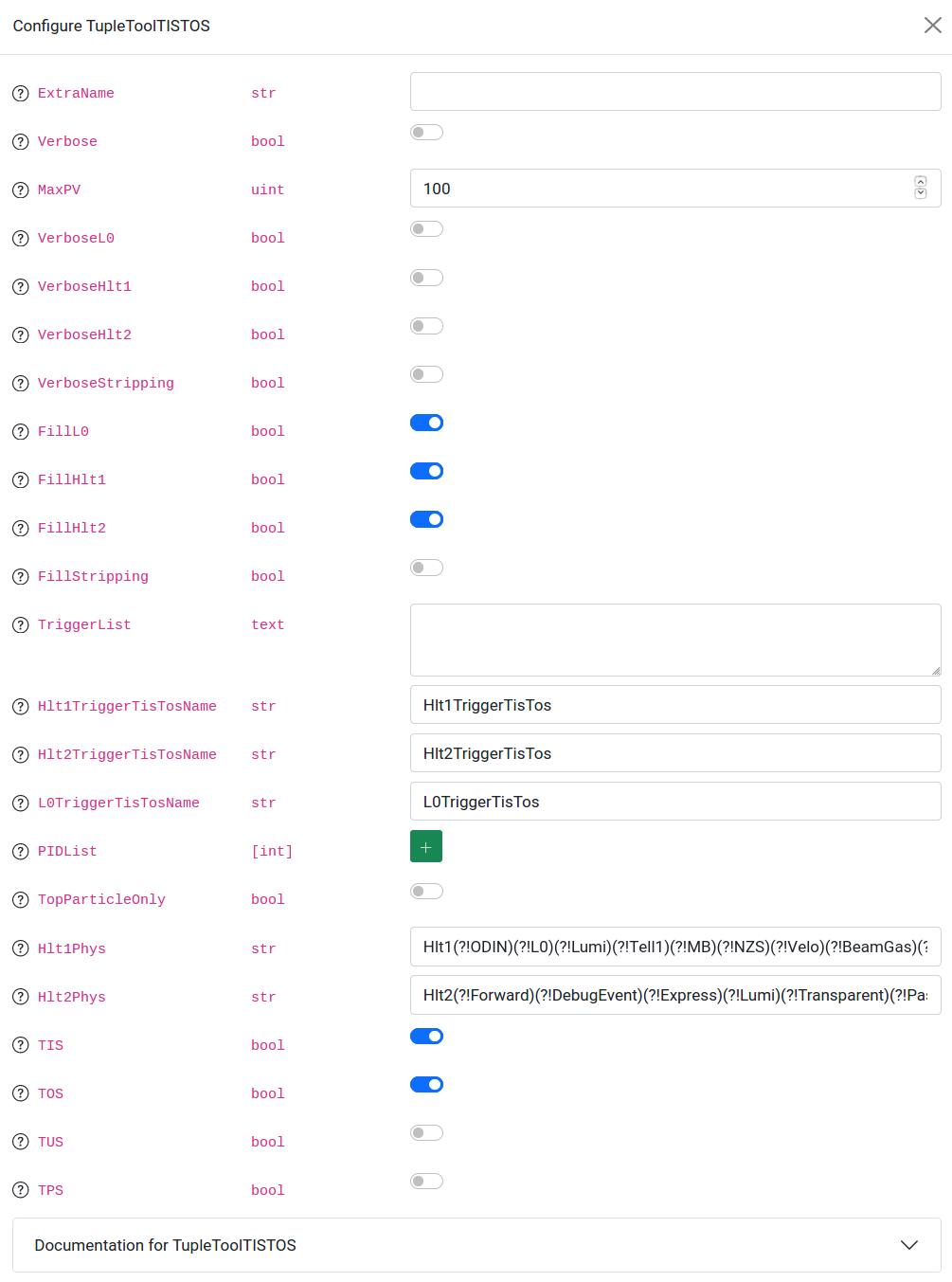} 
    \caption{Example of the configuration interface of a TupleTool within the \wizard{}, (in particular, \textit{TupleToolTISTOS} for saving trigger information), including links to relevant documentation at the bottom of the modal.}
    \label{fig:TupleTool_config}
\end{figure*}

Figure~\ref{fig:TupleTool_config} shows an example of the user interface for configuring TupleTools, with the particular example showing \textit{TupleToolTISTOS}, which saves trigger information to the Ntuple. It can be seen at the bottom how relevant information is provided. 

\subsection{Configuration output}
Figure~\ref{lst:dtt_yaml} shows an example of the output YAML file used to configure the DecayTreeTuple algorithm that was populated via configurations captured in Figs.~\ref{fig:node_tree} --~\ref{fig:TupleTool_config}, where the \texttt{tools, groups} and \texttt{branches} keys are shown specifying which TupleTools and therefore which information will be saved to the Ntuple. 
The top-level key \texttt{tools} contains a list of TupleTool configurations, from which the parsing functions create and configure TupleTool algorithms attached to the DecayTreeTuple itself, which will thus write either particle-level information about the decay or event-level information, depending on the class of the TupleTool.
The keys \texttt{branches} and \texttt{groups} themselves contain lists of dictionaries whose keys specify particles and have their own \texttt{tools} lists which are used similarly to attach TupleTool algorithms to the specified particle(s) in the decay tree.
Note that \texttt{groups} differs from \texttt{branches} in that it specifies multiple particles to be looped over and have identically configured TupleTool algorithms attached. 

\begin{figure*}[!htbp]
    \caption{Output of Btree.yaml, the data file used to configure the DecayTreeTuple algorithm.}
    \label{lst:dtt_yaml} 
    \begin{lstlisting}[language=yaml, frame=topline]    
inputs:
  - /Event/BhadronCompleteEvent/Phys/B2D0PiD2HHBeauty2CharmLine/Particles
descriptorTemplate: ${Bplus}[B+ -> ${D_0}(D~0 -> ${Kplus}K+ ${piminus}pi-)${piplus}pi+]CC
tools:
  - TupleToolKinematic:
      ExtraName: ''
      Verbose: false
      MaxPV: 100
      Transporter: ParticleTransporter:PUBLIC
  - TupleToolPid:
      ExtraName: ''
      Verbose: false
      MaxPV: 100
  - TupleToolANNPID:
      ExtraName: ''
      Verbose: false
      MaxPV: 100
      ANNPIDTunes:
        - MC12TuneV2
        - MC12TuneV3
        - MC12TuneV4
        - MC15TuneV1
      PIDTypes:
        - Electron
        - Muon
        - Pion
        - Kaon
        - Proton
        - Ghost
  - TupleToolGeometry:
      ExtraName: ''
      Verbose: false
      MaxPV: 100
      RefitPVs: false
      PVReFitter: LoKi::PVReFitter:PUBLIC
      FillMultiPV: false
  - TupleToolEventInfo:
      ExtraName: ''
      Verbose: false
      MaxPV: 100
branches:
  Bplus:
    particle: B+
    tools: []   
    \end{lstlisting}
\end{figure*}

\begin{figure*}[!htbp]
    \begin{lstlisting}[language=yaml, frame=bottomline]    
  D_0:
    particle: D~0
    tools: []
  Kplus:
    particle: K+
    tools: []
  piminus:
    particle: pi-
    tools: []
  piplus:
    particle: pi+
    tools: []
groups:
  Kplus,piminus:
    particles:
      - K+
      - pi-
    tools:
      - TupleToolTISTOS:
          ExtraName: ''
          Verbose: false
          MaxPV: 100
          VerboseL0: false
          VerboseHlt1: false
          VerboseHlt2: false
          VerboseStripping: false
          FillL0: true
          FillHlt1: true
          FillHlt2: true
          FillStripping: false
          TriggerList: []
          Hlt1TriggerTisTosName: Hlt1TriggerTisTos
          Hlt2TriggerTisTosName: Hlt2TriggerTisTos
          L0TriggerTisTosName: L0TriggerTisTos
          PIDList: []
          TopParticleOnly: false
          Hlt1Phys: >-
            Hlt1(?!ODIN)(?!L0)(?!Lumi)(?!Tell1)(?!MB)(?!NZS)(?!Velo)(?!BeamGas)(?!Incident).*Decision
          Hlt2Phys: >-
            Hlt2(?!Forward)(?!DebugEvent)(?!Express)(?!Lumi)(?!Transparent)(?!PassThrough).*Decision
          TIS: true
          TOS: true
          TUS: false
          TPS: false
name: DecayTreeTuple/Btree
    \end{lstlisting}
\end{figure*}

\subsection{Future developments}
It is planned to extend the current functionality of the \wizard{} by including the ability to create custom candidates from standard collections of LHCb particles. Another important planned addition is the ability to configure custom jet reconstruction. Ideally, support will be included for the full set of algorithms available in DaVinci for data analysis and event/candidate selection as resources allow.

As of Run 3, which started in 2022, the majority of the filtering and preselection of the data will be done in real time within the LHCb high-level trigger (HLT). In this architecture, the data will be fully reconstructed online and the final preselection algorithms will run in the HLT. Offline preselections will be feasible for a subset of the events. In both cases the output will have the same level of abstraction as the output of the stripping, allowing for a relatively simple adaptation of the Ntuple Wizard once the Run 3 data are made public.

\section{Request submission and execution}
\label{sec:prod}

Once the candidate(s) of interest, data set(s), and information to be saved in the Ntuple(s) are specified, and a name and email address have been provided for the production, the configuration files can be submitted to an LHCb Analysis Productions manager.

The Ntuple Wizard will be integrated with the CERN Open Data Portal, including an interface that allows users to make requests, check the status of their productions, and retrieve their output Ntuples all in one place. This will also provide a means to keep track of users through the CERN Open Data Portal database system. 

Requests for Ntuple creation are handled using the Analysis Productions package.
The files describing a new request are committed to a repository hosted on the CERN GitLab~\cite{git}, and a merge request is created once they are ready for review.
The Continuous Integration feature of GitLab is used to submit test productions to LHCbDIRAC~\cite{TDR1, TDR2}, which automatically processes a small fraction of the data when the remote repository is updated.

Once the request is submitted, it is handled by the LHCbDIRAC production system.
A production defines how a data set is to be processed, and LHCbDIRAC will launch and manage computing jobs until the data set is fully processed.
Productions are defined in `steps' that specify which application to run and which configuration files to read, and may be chained together such that the output of the first step is the input to the second, etc.
The \texttt{info.yaml} file produced by the \wizard{} defines one production per data set, each consisting of a single DaVinci step.

Within the production jobs, DaVinci is configured by functions defined in an external Python module according to the YAML files produced by the \wizard{}.
The data structure configured in Section~\ref{sec:alg} and displayed in Figure~\ref{lst:dtt_yaml} is traversed, and the configurable properties of the DecayTreeTuple algorithm are assigned the corresponding values.

After the Analysis Production jobs are complete, the produced Ntuples will be delivered to the CERN Open Data Portal for retrieval.
\section{Summary}
Providing public access to the large data sets at the LHC is a significant technical challenge, but it is becoming increasingly important for the longevity of high-energy physics in order to optimize acquired knowledge from the collected data. The volume and complexity of the data collected at LHCb make providing direct access to reconstructed (Level 3) data suitable for physics research difficult, motivating the design of the \wizard{}, where users can submit queries to obtain skimmed data samples (Ntuples) of the reconstructed data suitable for their physics interests. The \wizard{} is a web-based application that intuitively guides the user through specifying a query, from discovering a data set from a physics candidate (e.g. decay) of interest to configuring the information to be saved in the output Ntuple. The output of the \wizard{} is a pure data structure (YAML) format, which is to be submitted to an LHCb Analysis Productions manager so it can be parsed internally to provide the necessary Python scripts needed to configure the DaVinci application. The Ntuples will ultimately be delivered to the CERN Open Data Portal for retrieval.  


\section*{Appendices}
\appendix
\section{LHCb Data Organization}
\label{app:data_org}

Table~\ref{tab:runningyears} shows a list of running years, including corresponding collision energies and stripping versions available in the \wizard{}. 

\begin{table}[htbp!]
    \centering
    \caption{Table of running years, including collision energy ($\sqrt{s}$) and relevant stripping versions available in the \wizard{}.}    
    \begin{tabular}{|c|c|c|}
        \hline
        Running Year & $\sqrt{s}$ (TeV) & Stripping Versions \\
        \hline
        \multicolumn{3}{|c|}{Run 1} \\
        \hline
        2011 & 7 & s21r1, s21r1p1, s21r1p2 \\
        2012 & 8 & s21, s21r0p1, s21r0p2 \\
        \hline
        \multicolumn{3}{|c|}{Run 2} \\
        \hline
        2015 & 13 & s24r2 \\
        2016 & 13 & s28r2, s28r2p1 \\
        2017 & 13 & s29r2, s29r2p1, s29r2p2  \\
        2018 & 13 & s34, s34r0p1, s34r0p2 \\
        \hline
    \end{tabular}
    \label{tab:runningyears}
\end{table}

LHCb data streams come in two formats, Data Summary Tape (DST) files, which contain the full event information, and micro Data Summary Tape (MDST) files, which only contain the physics objects directly matched in at least one stripping selection. In MDST streams, the rest of the information in the events, apart from a few global event properties, is discarded. Table~\ref{tab:streams} shows a list of streams from Run 1 and Run 2, with the DST vs MDST format indicated in the stream name. 

\begin{table}[htbp!]
    \centering
    \caption{Table of data streams from Runs 1 and 2 available through the \wizard{}.}
    \begin{tabular}{|c|}
        \hline
         Stream \\
         \hline
         BHADRON.MDST  \\
         BHADRONCOMPLETEEVENT.DST \\
         CALIBRATION.DST  \\
         CHARM.MDST  \\
         CHARMCOMPLETEEVENT.DST  \\
         CHARMTOBESWUM.DST  \\
         DIMUON.DST  \\
         EW.DST  \\
         LEPTONIC.MDST  \\
         MINIBIAS.DST  \\
         PID.MDST  \\
         RADIATIVE.DST  \\
         SEMILEPTONIC.DST  \\
         \hline
    \end{tabular}
    \label{tab:streams}
\end{table}

\section{List of useful TupleTools}
\label{app:tupletools}
\subsection{Default TupleTools}
A set of TupleTools is included by default for all Ntuple configuration files produced by the \wizard{}. These tools can be removed by the user if desired, but are standard tools used in LHCb analyses, and are recommended to keep for physics analyses. Given the flexible data structure of the Ntuple, it is easy to produce a reduced data structure with a subset of variables at a later stage in data processing, while still maintaining the full set of variables in the original Ntuple. 
\begin{itemize}
    \item \textit{TupleToolANNPID} --- A tool used to add artificial neural network particle identification information about the physics candidate to the Ntuple.
    \item \textit{TupleToolEventInfo} ---  A tool used to add event and run information to the Ntuple.
    \item \textit{TupleToolGeometry} --- A tool used to add information about physics candidate geometry and event geometry to the Ntuple.
    \item \textit{TupleToolKinematic} --- A tool used to add kinematic information about the physics candidate to the Ntuple.
    \item \textit{TupleToolPid} --- A tool used to add particle identification information about the physics candidate to the Ntuple, with additional information than that in \textit{TupleToolANNPID}, including information about which PID detector subsystems were used in the probability calculations. 
\end{itemize}

\subsection{Other useful TupleTools}
\begin{itemize}
    \item \textit{TupleToolTISTOS} --- A tool that saves the trigger TIS/TOS (Trigger independent of Signal/Trigger on Signal) decisions for selected particles to the Ntuple.
    \item \textit{LoKi::Hybrid::TupleTool} --- A tool that allows the user to add LoKi variables or expressions of these variables known as LoKi functors to the Ntuple.
    
\end{itemize}

\section*{Acknowledgements}
%
%
\noindent We thank our colleagues at LHCb for providing the necessary data and software, and within the Data Processing \& Analysis (DPA) project for their incredibly valuable discussions. We additionally would like to thank Jose Marco Arias for systematic testing of the web interface. All authors acknowledge support from CERN.  In addition, C.A.A. and D.S.F. acknowledge support from the National Science Foundation under Award Number 2012926, and A.M. and S.N. acknowledge support from the DFG Grant NE 2185/1-1. 





\end{document}